\documentclass[aps,prl,showpacs,twocolumn,superscriptaddress]{revtex4}
\usepackage{amsmath, amssymb,mathrsfs}
\usepackage{graphicx}
\usepackage{color}

\DeclareMathOperator{\quabla}{\square} 
\DeclareMathOperator{\im}{Im}

\DeclareMathAlphabet{\mathjb}{OT1}{cmitt}{ub}{ui}
\DeclareMathAlphabet{\mathj}{OT1}{cmtt}{n}{ui}
\DeclareMathAlphabet{\mathpzc}{OT1}{pzc}{m}{ui}
\DeclareMathAlphabet{\mathe}{T1}{ectc}{m}{sl}
\renewcommand{\vec}[1]{\mathbf{#1}}

\begin{document}

\title{Propagation of Nonclassical Radiation through a Semiconductor
Slab}

\author{D.~Yu.~Vasylyev}
\affiliation{Arbeitsgruppe
Quantenoptik, Institut f\"ur Physik, Universit\"at Rostock, D-18051
Rostock, Germany}
 \affiliation{Bogolyubov Institute for Theoretical Physics, NAS of Ukraine, Metrolohichna 14-b, UA-03680 Kiev, Ukraine}

\author{W.~Vogel}
\affiliation{Arbeitsgruppe Quantenoptik, Institut f\"ur Physik,
Universit\"at Rostock, D-18051 Rostock, Germany}

\author{T.~Schmielau}
\affiliation{Centre for Electronic Devices and Materials, Sheffield Hallam University, United Kingdom}

\author{K.~Henneberger}
\affiliation{Arbeitsgruppe Festk\"orpertheorie, Institut f\"ur Physik,
Universit\"at Rostock, D-18051 Rostock, Germany}

\author{D.--G.~Welsch}
\affiliation{Theoretisch-Physikalisches Institut,
Friedrich-Schiller-Universit\"{a}t Jena, Max-Wien-Platz 1, D-07743
Jena, Germany}

\begin{abstract}
Based on a microscopic derivation of the emission spectra of a bulk semiconductor we arrive at a clear physical interpretation of the noise current operators in macroscopic quantum electrodynamics. This opens the possibility to study medium effects on nonclassical radiation propagating through an absorbing or amplifying semiconductor. As an example, the propagation of an incident squeezed vacuum is analyzed.
\end{abstract}


\maketitle

The progress of experimental techniques has rendered it possible to
almost completely measure the quantum statistics of
radiation--matter systems, so that correlation functions up to,
in principle, arbitrarily high orders can be detected
(for a review of the various methods, see, e.\,g., Ref.~\cite{we-vo-op}).
Within the framework of quantum optics, the theoretical study of the
interaction of light with complex material systems is
typically based on simplifying model systems, effective-Hamiltonian
schemes or related semi-phenomenological concepts rather
than rigorous microscopic calculations~\cite{vo-we}.

On the other hand, in many-particle quantum theory,
including semiconductor theory, much effort has been
spent on the development of methods for microscopically
describing complex material systems.
This includes the description of coherent optical
interactions by semiconductor Bloch equations
and semiconductor luminescence equations
as well as the development of nonequilibrium Green's function
methods~\cite{keldysh,HennebergerHaug}. In principle, these methods
lead to 
infinite hierarchies of correlations,
which
are usually treated by properly developed methods
of truncations and/or decorrelations.

It is well known that semiconductors can be used to
generate nonclassical radiation~\cite{Kim}.
In particular, the development of nano-structured systems
has opened new possibilities of the generation and
application of nonclassical radiation in integrated systems.
For example, the correlated emission of single photons
can
be demonstrated by using quantum dots~\cite{Michler}
and bound excitons in semiconductors~\cite
{Strauf}. Experiments with quantum wells~\cite{Hoyer}
and quantum dots~\cite{Stevenson,Young,Akopian} also
show the potential of semiconductors for the
generation of entangled photons, which are of interest
in quantum information processing.

In order to properly describe the generation and/or
propagation of nonclassical radiation through complex
material systems such as semiconductor slabs, one has
simultaneously to deal with both higher-order
radiation-field correlation functions and many-particle
quantum statistics of the material system.
Within the frame of macroscopic quantum electrodynamics (QED),
methods of describing the quantized electromagnetic field
in linearly responding (dispersing and absorbing/amplifying)
media have been developed, with special emphasis on
quantum-noise effects~\cite{Huttner,Gruner,Matloob97,
Scheel98,Raabe07,Raabe08}. Based on given constitutive 
relations of the material system,
field correlation functions of arbitrary order can be calculated
in this way. On the other hand, methods of many-particle
theory can be used to treat the radiation--matter
interaction within the frame of microscopic QED, thereby
calculating the relevant electromagnetic properties of
the material system. Such methods have been used to study the
propagation of radiation in semiconducting material,
including amplification and lasing~\cite{HennebergerKoch,Henneberger},
with the restriction to the intensity
and the emission spectrum of the radiation, because of the
elaborateness of the problem.

In the present paper we make the attempt to close
the gap between macroscopic
and microscopic QED, leading to a clear physical
interpretation of the noise-current density
introduced in macroscopic QED. In particular,
appropriate combination of the results of the two methods
offers the possibility to calculate arbitrary radiation-field
correlation functions, as we demonstrate by studying
the propagation of squeezed light through an absorbing/amplifying
semiconductor slab.

We begin with the operator of the vector
potential in Coulomb gauge, $\hat{\vec{A}}(\vec{r},t)$,
which obeys the inhomogeneous wave equation
\begin{equation}
   \quabla
   \hat{\vec{A}}(\vec{r},t)
    = -\mu_0  \,\hat{\vec{j}}(\vec{r},t)
\label{AGFeq}
\end{equation}
and the equal-time commutation rule
\begin{equation}
\left[\hat{\vec{A}}(\vec{r},t),
\frac{\partial}{\partial t}\,\hat{\vec{A}}(\vec{r}'\!,t)\right]
= \frac{i\hbar c}{\varepsilon_0 }\,
\overleftrightarrow{\delta}_{\!\!\mathrm{T}}(\vec{r}{-}{\vec{r}'}),
\label{AEcommut}
\end{equation}
where  $\overleftrightarrow{\delta}_{\!\!\mathrm{T}}(\vec{r})$
is the transverse tensorial delta function.
The operator of the transverse current density,
$\hat{\vec{j}}(\vec{r},t)$, can be decomposed into the
current density $\hat{\vec{j}}_{\rm med}(\vec{r},t)$ associated with
the medium under consideration and an auxiliary,
externally controlled
$c$-number current density $\vec{j}_{\rm ext}(\vec{r,t)}$
which is commonly set zero at the end of the calculations.

Starting from microscopic QED and employing 
Green's function (GF) techniques,
for
a
dielectric
medium in the steady state,
the
expectation value of the vector potential
(in the Fourier domain)
reads as
\begin{equation}
\langle\hat{A}(\vec{r},\omega)\rangle
   {=} -\mu_0\int d^3\vec r'\,
   \overleftrightarrow{D}^{{\rm ret}}(\vec{r},\vec{r}',\omega)
\cdot
\vec{ j}_\mathrm{ext}(\vec{r'},\omega)
,
\label{solut}
\end{equation}
where, in linear approximation,
the inverse of the retarded photon GF
$\overleftrightarrow{D}^{{\rm ret}}(\vec{r},\vec{r}',\omega)$ is given by
\begin{equation}
   \overleftrightarrow{D}^{{\rm ret},-1}(\vec{r},\vec{r}',\omega) = \quabla
   \delta(\vec{r}{-}\vec{r}')
   - \overleftrightarrow{P}^{\rm ret}(\vec{r},\vec{r}',\omega)\label{Dinv}
,
\end{equation}
with the
retarded polarization tensor
$\overleftrightarrow{P}^{\rm ret}(\vec{r},\vec{r}',\omega)$
being related to the
complex (dielectric)
susceptibility tensor $\overleftrightarrow{\chi}(\vec{r},\vec{r}',\omega)$
$\!=$ $\overleftrightarrow{\chi}'(\vec{r},\vec{r}',\omega)$
$\!+$ $\!i\overleftrightarrow{\chi}''(\vec{r},\vec{r}',\omega)$
of the medium according to
\begin{equation}
 \overleftrightarrow{P}^{\rm ret}(\vec{r},\vec{r}',\omega) =
- \frac{\omega^2}{c^2}\,\overleftrightarrow{\chi}(\vec{r},\vec{r}',\omega)
.
\label{chi}
\end{equation}
The Keldysh components of the
photon GF,
\begin{multline}
\overleftrightarrow{D}^>
   (\vec{r},\vec{r}'\!,t{-}t')=
\overleftrightarrow{D}^<
   (\vec{r}',\vec{r},t'-t)\\
   = \frac{1}{i\hbar}\left[\langle
   \hat 
{\vec{A}}
   (\vec{r},t)\hat 
{\vec{A}}
   (\vec{r}',t')\rangle-\langle
   \hat 
{\vec{A}}
  (\vec{r},t)\rangle\langle \hat
{\vec{A}}
  (\vec{r}',t')\rangle\right]
,
\label{Dgtrless1}
\end{multline}
describe 
field--field fluctuations. Solving the Dyson equation
for these components, one obtains the optical theorem
\begin{multline}
   D_{ij}^\gtrless(\vec{r},\vec{r}',\omega)
    = \int d^3\vec r_1 \int  d^3\vec r_2\, 
    D_{ik}^{\rm ret}(\vec{r},\vec{r}_1,\omega)
\\
   \times\, P_{kl}^\gtrless(\vec{r}_1,\vec{r}_2,\omega)
   \,D_{lj}^{\rm adv}(\vec{r}_2,\vec{r}',\omega),
\label{Dgtrless}
\end{multline}
where $P_{kl}^\gtrless$ are the corresponding Keldysh components of the
polarization 
tensor,
and
summation over repeated indices is understood.
Within the framework of
macroscopic QED,
Eq.~(\ref{AGFeq}) is regarded as being the operator-valued
inhomogeneous wave equation that corresponds to the Maxwell equations
of the transverse part of the macroscopic electromagnetic field
in a linear dielectric medium. Hence, $\hat{\vec{j}}_\mathrm{med}(\vec{r},\omega)$
must have the form
\begin{equation}
 \vec{\hat j}_{\rm med}(\vec r,\omega){=}{-}\varepsilon_0\omega^2\!\int d^3\vec r'\,
\overleftrightarrow{\chi}(\vec r,\vec r',\omega)
{\cdot}
\vec{\hat A}(\vec r',\omega)
+ \hat{\vec{j}}_\mathrm{N}(\vec r,\omega)
,
\end{equation}
and in
place of Eq.~(\ref{solut}) we obtain
the operator-valued equation
\begin{multline}
\vec{\hat A}(\vec{r},\omega)
\\
{=}{-}\mu_0\int d^3\vec r'\,
\overleftrightarrow{D}^{\rm ret}(\vec r,\vec r',\omega)
{\cdot}\!\left[
\vec{\hat j}_{\rm N}(\vec r',\omega)
{+} \vec{j}_\mathrm{ext}(\vec r',\omega)
\right]
,
\label{Asolut}
\end{multline}
where the
noise current density operator $\vec{\hat j}_{\rm N}(\vec r,\omega)$ obeys 
the commutation relation~\cite{Raabe07}
 \begin{equation}
      \bigl[\vec {\hat j}_{\rm N}(\vec r,\omega),
      \vec{\hat  j}_{\rm N}^{\dagger}(\vec r'\mspace{-3mu},\omega')\bigr]
      = 2\varepsilon_0
      \hbar \omega^2
\overleftrightarrow{\chi}''(\vec r,\vec r'\omega)
       \delta(\omega{-}\omega')
\label{jNcommut}
   \end{equation}
[$\vec{\hat  j}_{\rm N}(\vec r,-\omega)$ $\!=$
$\!\vec{\hat  j}_{\rm N}^\dagger(\vec r,\omega)$],   
   which ensures the validity of 
Eq.~(\ref{AEcommut}). 

With the analog of the optical theorem~(\ref{Dgtrless}) in macroscopic QED, the relation 
\begin{equation}
\begin{split}
 \overleftrightarrow{P}^>(\vec r,\vec r',t{-}t')&=\overleftrightarrow{P}^<(\vec r',\vec r,t'{-}t)\\
&= \frac{\mu_0}{i\hbar}\bigl\langle
\hat{\vec{j}}_\mathrm{N}
(\vec r,t)
\hat{\vec{j}}_\mathrm{N}
(\vec r',t')\bigr\rangle
\label{Pjj}
\end{split}
\end{equation}
can be derived.
Moreover, using
Green's function techniques, a Bethe-Salpeter equation between the polarization function and  the correlation function of the medium current density  can be derived \cite{Schaefer}.
In this way, the fluctuation of the noise 
current density in 
macroscopic QED 
can be
directly related to the
fluctuation of the  
microscopically well-defined and observable medium 
current density.
 
In the following we will deal with the propagation of 
TE-polarized radiation
along the $x$ axis, through a semiconductor slab of 
thickness $L$ which is infinitely extended in the 
$y$-$z$--plane. For simplicity, we assume
that the electric field is polarized along the 
$y$-axis  $\vec{\hat A}{=}(0,\hat A,0)$. Then, neglecting 
spatial dispersion, the complex refractive index 
inside the medium, $n=n'{+}in''$, is obtained from
$n^2(\omega)=1 + \chi (\omega) $.

Using 
Green's function technique, 
the spontaneous
emission of the slab 
is studies in Ref.~\cite{HennebergerKoch,Henneberger}. 
A generalization of 
the
results including
spatial dispersion and providing exact relations between 
(spontaneous and stimulated) emission and linear absorption 
can be found in Ref.~\cite{arxiv}. The 
intensity of the (spontaneously) emitted radiation 
can be given by
 \begin{equation}
         I(\omega) = b(\omega) 
P
(\omega) {\hat D}_0 (\omega)\, ,
 \label{intensity2}
 \end{equation}
where ${\hat D}_0$ is the vacuum-induced contribution 
to the photon spectral function, 
$P = 2i \im P^{\rm ret}$
is, 
according to (\ref{chi}), related to 
$\chi''$
associated with linear absorption/amplification, 
and  $b$ is defined as the ratio between the 
recombination rate 
$P^<$
of electron--hole pairs 
and 
$P$.
The 
function
$b(\omega)$ characterizes globally 
(i.\,e., inside and outside the slab) 
the emitted 
radiation
and, as
such, $b$ is accessible to direct observation in experiments.
It generalizes Planck's formula for the black body radiation to 
nonequilibrium radiation of an excited medium 
in the steady state.

On the basis of Eq.~(\ref{Asolut}),
input--output relations 
can be derived, by introducing
bosonic quasiparticle annihilation operators 
(normalization $\mathcal{N}_\pm$)
  \begin{equation}
     {\hat  c}_{\pm}(\omega){=} \mathcal{N}_\pm(\omega)\!\!\!
     \int_{-L/2}^{L/2}
     dx'\left(e^{i n\omega
     x'/c}\pm\mspace{-4mu} e^{-i n\omega
     x'/c}\right)\,\,{\!\!\hat { j}}_{\rm N}(x',\omega)
\label{cplusminus}
  \end{equation}
expressed 
in terms of the noise current 
density operator and associated with the slab--radiation 
excitations~\cite{Gruner,Knoell99}.
On the other hand, the
poles of $D^{\rm ret}$ yield the 
polaritonic
dispersion relations for
the slab,
which
allows one to relate 
the operators $\hat{c}_\pm$ 
to
the polaritonic annihilation 
operators~\cite{HuttnerBaumberg}.
Assuming that the input 
field is
in the vacuum state, 
the intensity 
of the output field
is obtained
in just the same form as in Eq.~(\ref{intensity2}), 
by identifying $\langle  {\hat c}^\dagger(\omega)
{\hat c}(\omega) \rangle \equiv b(\omega)$,
where ${\hat c}(\omega)={\hat c}_+(\omega) + {\hat c}_-(\omega)$.

Steadily
excited semiconductors in quasi-equilibrium
are of particular interest,
such as
exciton gases generated at low up to moderate excitation and
light-emitting diodes working at high excitation.
For quasi-equilibrium, due to the Kubo-Martin-Schwinger 
condition~\cite {KMS57}, the function $b(\omega)$ develops 
into a Bose distribution $\,b (\omega) =
\left( {\rm exp} \left[\beta(\hbar \omega - \mu)\right] 
- 1 \right)^{-1}\,$.  The chemical potential $ \mu $ starts 
at $ \mu = 0 $ for complete thermal equilibrium
and characterizes the degree of excitation beyond the 
thermal one for $ \mu > 0 $. The crossover from absorption 
($\chi''>0$) to gain ($\chi''<0$) appears at $ \hbar \omega = \mu $.
By expanding the product $b(\omega) \chi''(\omega)$ 
in Eq.~(\ref{intensity2}) at $\hbar \omega = \mu $,
it is seen that the spontaneous emission remains finite  
at the crossover, it is given by the slope of the absorption 
function $\chi''$. Since both $\chi''$ and $b$ switch their signs, 
the spontaneous emission stays positive in the whole 
frequency region, as it should be. Thus we obtain a unified 
description of absorption and amplification in
semiconductors.

In order to demonstrate the usefulness of the established 
interrelation of microscopic and macroscopic QED,
we consider the propagation of squeezed light through 
a semiconductor slab.
Consider the squeezed vacuum state 
   \begin{equation}
      |\psi\rangle_{\mathrm{sv}}
      =\hat S|\psi\rangle_{\mathrm{v}},\label{psi}
   \end{equation}
where the squeeze operator,
   \begin{equation}
      \hat S=\exp\left\{ \int_0^{\infty}d\omega
      \left[\xi^{\ast}\hat a
      (\omega_0{+}\omega)\hat a (\omega_0{-}\omega)-\text{h.c.}\right]
       \right\}
   \end{equation}
acts on the multimode vacuum state $|\psi\rangle_{\mathrm{v}}$. 
The squeeze parameter is $\xi=|\xi|e^{i\phi_{\xi}}$, and
$\hat a(\omega)$ are the photonic annihilation operators
of the incident field on the left-hand side of the semiconductor 
slab. The input field on the right-hand side is assumed to 
be in the vacuum state.

\begin{figure}
\includegraphics[width=1\linewidth,clip=]{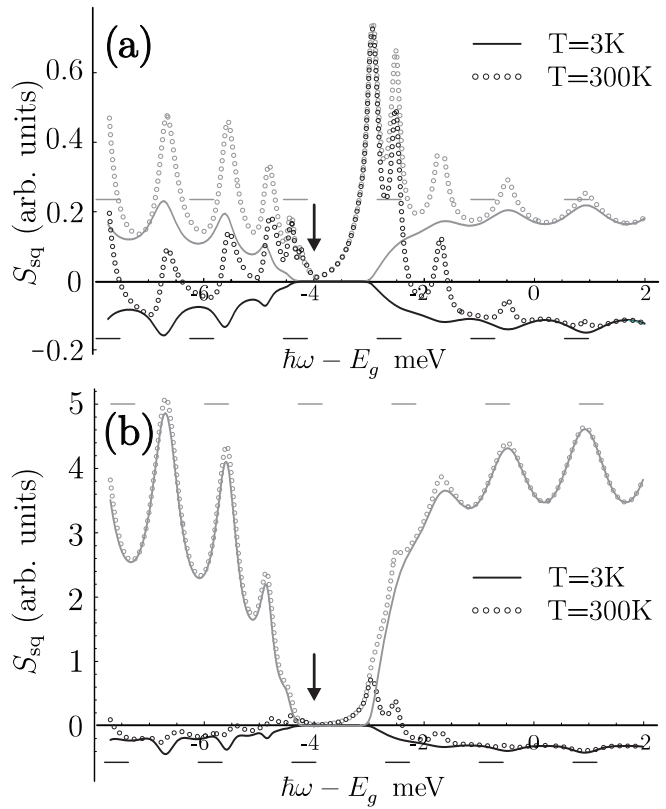}
\caption{\label{sq} Output squeezing spectrum for a
GaAs slab ($L=25\mu m$) near the $1s$-excitonic resonance (indicated by an arrow) of bandwidth
$=0.2$ meV. The squeezing spectrum is shown for two
temperatures and for $|\xi|=0.2$ (a), $1.2$ (b). The dashed lines indicate the
maximum and minimum noise level of the squeezed input field.}
\end{figure}%
\begin{figure}
\includegraphics[width=1\linewidth,clip=]{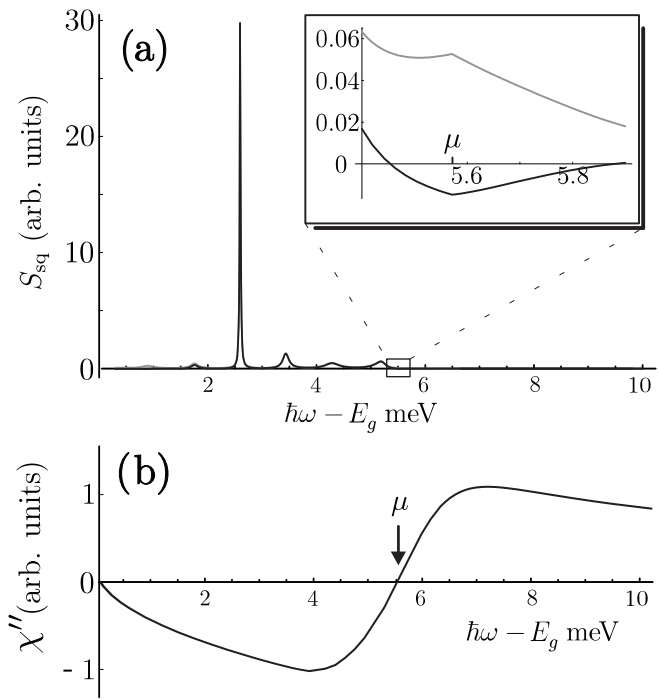}
\caption{\label{sq1}
Squeezing spectrum after transmission of a
GaAs slab at temperature T=3K, pumped to a
carrier density of $n=3\cdot10^{16}cm^{-3}$ (a) and the corresponding
imaginary part of susceptibility (b).
The input (dashed) and transmitted (solid) squeezing spectra are
shown  for $|\xi|=0.2$.  An arrow indicates the
chemical potential $\mu$ and the inset shows its vicinity.
}
\end{figure}%
Using  Eqs~(\ref{Asolut}), (\ref{intensity2}), and (\ref{psi}), 
the normally ordered squeezing spectrum $S_{\rm sq}
=\int d\omega' \omega_0^2\langle 
:\hat A(x,\omega)\hat A(x',\omega'):\rangle_{\mathrm{sv}}$ is 
derived for $x{=}x'{=}L$ as
   \begin{multline}
      S_{\rm sq}=I(\omega)+\frac{\hbar\omega_0}
      {2\pi \varepsilon_0 c}\Bigl\{
      |\mathcal{T}(\omega)|^2\sinh^2|\xi|
      \\
      + e^{{-}2i \omega_0L/c{-}i\phi_{\xi}}
      \left[\mathcal{T}^{\ast}\mspace{-3mu}(\omega)\right]^2
      \cosh\mspace{-3mu}|\xi|
      \sinh\mspace{-3mu}|\xi|{+}\text{c.c.}\Bigr\}.\label{Ssq}
   \end{multline}
Here  we have assumed that the detector is placed on the 
right-hand side of the slab. The total transmission coefficient 
of the slab,  $\mathcal{T}(\omega)$, reads in terms of the 
internal reflectivity $r(\omega){=}[1{-}n(\omega)]
/[1{+}n(\omega)]e^{i\omega n(\omega)L/c}$ as \cite{Matloob}
\begin{equation}
\mathcal{T}(\omega){=}\frac{4n(\omega)}{[1{+}n(\omega)]^2\left[1{-}r^2(\omega)\right]} \, e^{i\omega (n(\omega){-}1) L/c}.
\end{equation}
From Eqs~(\ref{intensity2}) and (\ref{Ssq}) it is evident 
that the behavior of the function $S_{\rm sq}$ strongly 
depends on the behavior of the denominator $1-r^2(\omega)$. 

So,
the squeezing spectrum 
shows an
oscillating behavior 
with the maxima at the frequencies $\omega_s=\frac{2\pi c}
{L n'(\omega)}s$ with integer $s$ (Fabry--Perot resonances). 

In Fig.~\ref{sq} we show the
influence of a single excitonic resonance at $E_x{=}\hbar\omega_0$ on
squeezed light propagating through a semiconductor slab, for different
values of the squeezing strength $|\xi|$
where the
Lorentz oscillator model 
has been
used to simulate excitonic absorption \cite{HaugKoch}.
The input (squeezed white noise) spectrum is depicted with dashed
lines. The upper (gray) and lower (black) curves represent 
the maximum and the minimum noise level of the squeezing 
spectrum of the radiation after transmission through the 
slab. 
For high temperatures, the emission spectrum $I(\omega)$ 
plays a more pronounced role and it leads to a significant 
decrease of the nonclassical properties of the input field.

The squeezing spectrum for a highly excited semiconductor is 
shown in Fig.~\ref{sq1}(a) for $|\xi|=0.2$. The susceptibility 
function  has been calculated in the effective 
pair-equation approximation~\cite{HaugKoch}
and is given in Fig.~\ref{sq1}(b). In the output field, the squeezing
is nearly destroyed, due to the dominance of the incoherent 
emission in the gain region.
Minor squeezing is only preserved in the vicinity of the 
crossover at $\hbar\omega=\mu$, where $\chi''=0$.

In conclusion, we have 
studied
the relation between 
microscopic and macroscopic 
QED,
leading to a clear physical 
interpretation of some basic quantities. By combining the 
input--output formalism of 
macroscopic QED with results of microscopic 
QED,
the treatment of the propagation of nonclassical light 
in complex media becomes possible. As an example, 
we have studied squeezed-light propagation through 
a semiconductor slab.

This work was supported by the Deutsche 
Forschungsgemeinschaft, Sonderforschungsbereich 652.


\begin{thebibliography}{1}


\bibitem{we-vo-op} D.-G. Welsch, W. Vogel, and T. Opatrny, 
 in: {\it Progress in Optics}, Ed. E. Wolf, Vol. XXXIX, pp. 63 (1999).

\bibitem{vo-we} W. Vogel and D.--G. Welsch, {\it Quantum Optics}  (Wiley-VCH, Weinheim, 2006).


\bibitem{keldysh} L. V. Keldysh, Zh. eksper. teor. fiz. {\bf 47}, 1515 (1964).

\bibitem{HennebergerHaug} K. Henneberger and H. Haug, Phys. Rev. B {\bf 38}, 9759 (1988).

\bibitem{Kim} J. Kim, S. Somani and Y. Yamomoto, {\it Nonclassical light from semiconductor lasers and
LEDs} (Springer, Berlin, 2001).

 \bibitem{Michler} P. Michler, A. Kiraz, C. Becher, W. Schoenfeld, P. Petroff,
  L. Zang, E. Hu, A. Imamoglu, {\em Science} {\bf 290}, 2282 (2000).

\bibitem{Strauf} S. Strauf, P. Michler, M. Klude, D. Hommel, G. Bacher,
  A. Forchel, Phys. Rev. Lett. {\bf 89}, 177403 (2002).

\bibitem{Hoyer} W. Hoyer, M. Kira, S.W. Koch, H. Stolz, S. Mosor, J.
  Sweet, C. Ell, G. Khitrova, and H.M. Gibbs, Phys. Rev. Lett. {\bf 93},
  067401 (2004).

\bibitem{Stevenson} R.M. Stevenson, R.J. Young, P. Atkinson, K. Cooper,
  D.A. Ritchie, and A.J. Shields, Nature {\bf 439}, 179 (2006).

\bibitem{Young} P.J. Young,  R.M. Stevenson, P. Atkinson, K. Cooper,
  D.A. Ritchie, and A.J. Shields, New J. Phys. {\bf 8}, 29 (2006).

\bibitem{Akopian} N. Akopian, N.H. Lindner, E. Poem, Y. Berlatzky, J. Avron,
  D. Gershoni, B.D. Gerardot and P.M. Petroff, Phys. Rev. Lett. {\bf 96}, 130501 (2006).

\bibitem{Huttner} B. Huttner and S. M. Barnett,
Phys. Rev. A {\bf 46}, 4306 (1992).

\bibitem{Gruner} T. Gruner and D.--G. Welsch,
Phys. Rev. A {\bf 53}, 1818 (1996).

\bibitem{Matloob97} R. Matloob, R. Loudon, M. Artoni, S. M. Barnett,
and J. Jeffers,
Phys. Rev. A \textbf{55}, 1623 (1997).

\bibitem{Scheel98} S. Scheel, L. Kn\"{o}ll, and D.-G. Welsch,
Phys. Rev. A \textbf{58}, 700 (1998).

\bibitem{Raabe07} C. Raabe, S. Scheel, and D.-G. Welsch,
Phys. Rev. A \textbf{75}, 053813 (2007).

\bibitem{Raabe08} C. Raabe  and D.-G. Welsch,
Eur. Phys. J., in press (arXiv:0710.2867v1 [quant-ph]).

\bibitem{Schaefer}  W. Sch\"{a}fer and M. Wegener {\it Semiconductor Optics and Transport Phenomena}, (Springer, Berlin, 2002).

\bibitem{HennebergerKoch} K. Henneberger and S.W. Koch,
Phys. Rev. Lett. {\bf 76}, 1820 (1996).

\bibitem{Henneberger} K. Henneberger and S.W. Koch in {\it Quantum Theory of the Optical and Electronic
Properties of Semiconductors}, ed.\  S. W. Koch (World Scientific, Singapore, 1996).

\bibitem{Knoell99} L. Kn\"oll, S. Scheel, E. Schmidt, D.-G. Welsch,
and A.V. Chizhov, Phys. Rev. A \textbf{59}, 4716 (1999). 

\bibitem{arxiv} K. Henneberger, arXiv:0710.5686 [condmat.str-el].

\bibitem{HuttnerBaumberg} B. Huttner, J.J. Baumberg and S.M. Barnett,  Europhys. Lett. {\bf 16}, 177 (1991).

\bibitem{KMS57} R. Kubo, J. Phys. Soc. Jpn. {\bf 12}, 570 (1966); P.C. Martin and J. Schwinger, Phys.\ Rev. {115},  1432 (1959).

\bibitem{Matloob} R. Matloob, R. Loudon, S.M. Barnett, and J. Jeffers,
Phys.\ Rev.\ A {\bf 52}, 4823 (1995).

\bibitem{HaugKoch} H. Haug and S.W. Koch,  {\it Quantum theory of the optical and electronic properties  of semiconductors} (World Scientific, Singapore, 2005).

\end{thebibliography}
\end{document}